\documentclass[structabstract]{aa}
\usepackage{float}
\usepackage{lineno}
\usepackage{natbib}
\bibpunct{(}{)}{;}{a}{}{,}
\usepackage{graphicx}
\usepackage{epsfig}
\usepackage{amsmath}
\usepackage[varg]{txfonts}
\usepackage{lscape}
\usepackage{longtable}
\usepackage{txfonts}
\usepackage{color}

 \newcommand{\refA}{\color{black}}
 
\newcommand{\multilineL}[1]{\begin{tabular}[b]{@{}l@{}}#1\end{tabular}}

\makeatletter

\newcommand{\Rmnum}[1]{\expandafter\@slowromancap\romannumeral #1@}
\makeatother
\begin{document}
   \title{Full-disk nonlinear force-free field extrapolation of SDO/HMI and SOLIS/VSM magnetograms}
   \author{Tilaye Tadesse\inst{1, 2}, T. Wiegelmann\inst{3}, B.
   Inhester\inst{3}, P. MacNeice\inst{4}, A. Pevtsov\inst{5}, \and X. Sun\inst{6}
          }
   \institute{Department of Physics, Drexel University, Philadelphia, PA 19104-2875, USA \\
              \email{tasfaw@einstein.physics.drexel.edu}
    \and Addis Ababa University, Institute of Geophysics, Space Science, and Astronomy, Po.Box 1176, Addis Ababa, Ethiopia\\
      \email{tilaye.tadesse@gmail.com}
\and Max-Planck-Institut f\"ur Sonnensystemforschung,
              Max-Planck-Strasse 2, 37191 Katlenburg-Lindau, Germany
   \and NASA/ GSFC, Greenbelt, MD, U.S.A.
   \and National Solar Observatory, Sunspot, NM 88349, U.S.A.
    \and W.W. Hansen Experimental Physics Laboratory, Stanford
University, Stanford, CA 94305, USA
      }
  \abstract
{The magnetic field configuration is essential to understand solar explosive phenomena such as flares and coronal 
mass ejections. To overcome the unavailability of coronal magnetic field measurements, photospheric magnetic field vector 
data can be used to reconstruct the coronal field. A complication of this approach is that the measured photospheric 
magnetic field is not force-free and that one has to apply a preprocessing routine in order to achieve boundary 
conditions suitable for the force-free modelling. Furthermore the nonlinear force-free extrapolation code should 
take uncertainties into account in the photospheric field data  which occur due to noise, incomplete inversions or 
azimuth ambiguity removing techniques.
}
{Extrapolation codes in cartesian geometry for modelling the magnetic field in the corona do not take the curvature of 
the Sun's surface into account and can only be applied to relatively small areas, e.g., a single active region. Here we 
apply a method for nonlinear force-free coronal magnetic field modelling and preprocessing of photospheric vector magnetograms 
in spherical geometry using the optimization procedure to full disk vector magnetograms. We compare the analysis of the
photospheric magnetic field and subsequent force-free modeling based on full-disk vector maps from Helioseismic and Magnetic 
Imager (HMI) on board solar dynamics observatory (SDO)  and Vector Spectromagnetograph (VSM) of the Synoptic Optical Long-term 
Investigations of the Sun (SOLIS).
}
{We use Helioseismic and Magnetic Imager and Vector Spectromagnetograph photospheric magnetic field measurements to model 
the force-free coronal field above multiple solar active regions, assuming magnetic forces to dominate. We solve the nonlinear 
force-free field equations by minimizing a functional in spherical coordinates over a full disk excluding the poles. After 
searching for the optimum modeling parameters for the particular data sets, we compare the resulting nonlinear force-free 
model fields. We compare quantities like the total magnetic energy  content and free magnetic energy , the longitudinal distribution 
of the magnetic pressure and surface electric current density using our spherical geometry extrapolation code.
}
{The magnetic field lines obtained from nonlinear force-free extrapolation based on Helioseismic and Magnetic Imager and Vector 
Spectromagnetograph data have good agreement. However, the nonlinear force-free extrapolation based on Helioseismic and Magnetic 
Imager data have more contents of total magnetic energy, free magnetic energy, the longitudinal distribution of the magnetic pressure 
and surface electric current density compared to the one from Vector Spectromagnetograph data.
}

\titlerunning{Full-disk NLFFF extrapolation of SDO/HMI and SOLIS/VSM data}
\authorrunning {T. Tadesse \textit{et al}.}
   \maketitle
\section{Introduction}
The magnetic field’s configuration is essential for us to understand solar explosive phenomena such as flares 
and coronal mass ejections. The corona has been the subject of extensive modeling for decades, but these efforts 
have been hampered by our limited ability to determine the corona's three-dimensional structure \citep{Schrijver:2011,Sandman:2011}. 
Since the corona is optically thin, direct measurements of these magnetic fields are very difficult to implement, 
and the present observations for the magnetic fields based on the spectropolarimetric method (the Zeeman and the 
Hanle effects) are limited to low layers of solar atmosphere (photosphere and chromosphere).  The problem of measuring 
the coronal field and its embedded electrical currents thus leads us to use numerical modelling to infer the field 
strength in the higher layers of the solar atmosphere from the measured photospheric field. Due to the low value of 
the plasma $\beta$ (the ratio of gas pressure to magnetic pressure), the solar corona is magnetically dominated \citep{Gary}. 
To describe the equilibrium structure of the static coronal magnetic field when non-magnetic forces are negligible, the force-free
assumption is appropriate:
\begin{equation}
   (\nabla \times\textbf{B})\times\textbf{B}=0 \label{one}
\end{equation}
\begin{equation}
    \nabla \cdot\textbf{B}=0 \label{two}
 \end{equation}
 subject to the boundary condition
\begin{equation}
    \textbf{B}=\textbf{B}_{\textrm{obs}} \quad \mbox{on photosphere} \label{three}
 \end{equation}
where $\textbf{B}$ is the magnetic field and $\textbf{B}_{\textrm{obs}}$ is measured vector field on the
photosphere. Equation~(\ref{one}) states that the Lorentz force vanishes (as a consequence of
$\textbf{J}\parallel \textbf{B}$, where $\textbf{J}$ is the electric current density) and Equation~(\ref{two})
describes the absence of magnetic monopoles. Based on the above assumption, the coronal magnetic
field is modelled with nonlinear force-free field (NLFFF) extrapolation
\citep{Inhester06,valori05,Wiegelmann04,Wheatland04,Wheatland:2009,tilaye09,Wheatland:2011,Amari:2010,Wiegelmann:2012,Jiang:2012}.
From a mathematical point of view appropriate boundary condition for force-free modeling are the vertical
magnetic field $B_{n}$ and the vertical current $J_{n}$ prescribed only for one polarity of $B_{n}$
\citep{Amari97,Amari99,Amari}. A direct use of these boundary conditions is implemented in Grad-Rubin codes
\citep{Amari99}. \citet{Wheatland:2009} and \citet{Wheatland:2011} implemented the use of $B^{+}_{n}$ and $B^{-}_{n}$
solution together with an error approximation to derive consistent solutions.
Using the three components of $B$ as boundary condition requires consistent magnetograms, as outlined in
\citet{Aly89}. We use preprocessing and relaxation of the boundary condition to derive these consistent
data on the boundary.

As an alternative to real measurement, nonlinear force-free field (NLFFF) models are thought to be viable tools for 
investigating the structure, dynamics, and evolution of the coronae of solar active regions. It has been found that
NLFFF models are successful in application to analytic test cases \citep{Schrijver06,Metcalf:2008}, but
they are less successful in application to real solar data. {\refA{However, NLFFF models have been adopted to study 
various magnetic field structures and properties in the solar atmosphere. For instance, \citet{Regnier:2002,Regnier:2004,Canou:2009,Canou:2010,Guo:2010,Valori:2012} 
have substantially studied various magnetic field structures and properties using their respective NLFFF model codes.}} 
Different NLFFF models have been found to have markedly different field line configurations and to provide widely varying estimates
of the magnetic free energy in the coronal volume, when applied to solar data \citep{DeRosa}. The main
reasons for that problem are (1) the forces acting on the field within the photosphere, (2) the uncertainties
on vector-field measurements, particularly on the transverse component, and (3) the large domain that
needs to be modelled to capture the connections of an active region to its surroundings\citep{Tilaye:2010,Tilaye:2012}. 
In this study, we have considered those three points explicitly into account. However, caution must still be needed while
assessing results from this modeling. This is because many aspects of the specific approach to modeling
used in this work, such as the use of preprocessed boundary data, the missing boundary data, and the
departure of the model fields from the observed boundary fields may influence the results. 

In this work, we use full-disk SDO/HMI and SOLIS/VSM photospheric magnetic field measurements to model 
the NLFFF coronal field above multiple solar active regions. Comparison of vector magnetograms for 
one particular active region observed with two different instruments from SOLIS and HMI and their corresponding 
force-free models have been studied by \citet[][submitted to AJ]{Thalmann:2012} in Cartesian coordinates. We use a larger 
computational domain which accommodates most of the connectivity within the coronal region. We use a spherical 
version of the optimization procedure that has been implemented in \citet{Tilaye:2010}. We compare quantities 
like the total magnetic energy  content and free magnetic energy  and the longitudinal distribution of the magnetic 
pressure in the HMI and VSM-based model volumes in spherical geometry. We relate the appearing differences to the 
photospheric quantities such as the magnetic fluxes and electric currents but also show the extent of agreement 
of NLFFF extrapolations from different data sources.
\section{Instrumentation and data set}
\subsection{Solar Dynamics Observatory(SDO) - Helioseismic and Magnetic Imager(HMI)}
The Helioseismic and Magnetic Imager \citep[HMI;][]{ Schou:2012} is part of the Solar Dynamics Observatory (SDO)
and observes the full Sun at six wavelengths and full Stokes profile in the Fe {\Rmnum{1}} 617.3 nm spectral line. HMI 
consists of a refracting telescope, a polarization selector, an image stabilization system, a narrow band tunable filter 
and two 4096 pixel CCD cameras with mechanical shutters and control electronics. Photospheric line-of-sight LOS and vector 
magnetograms are retrieved from filtergrams with a plate scale of 0.5 arc-second. From filtergrams averaged over about 
ten minutes, Stokes parameters  are derived and inverted using the Milne-Eddington (ME) inversion algorithm of \citet{Borrero:2011} 
(the filling  factor is held at unity). Within automatically identified regions of strong magnetic fluxes \citep{Turmon:2010}, 
the full disk inversion data are from the second HMI vector data release (JSOC data series hmi.ME\_720s\_e15w1332). The 180-degree 
azimuthal ambiguity in the strong field region is resolved using the Minimum Energy Algorithm \citep{Metcalf:1994,Metcalf:2006,Leka:2009}, 
taken from the AR patches in the second release also (data series hmi.B\_720s\_e15w1332\_cutout). For the weak field region where 
noise dominates, we adopt a radial-acute angle method to resolve the azimuthal ambiguity. The weak field region is defined as 
where field strength is below 200 G at disk center, 400 G on the limb, and varies linearly in between. The noise level is 
$\approx$ 10G and $\approx$ 100G for the longitudinal and transverse magnetic field, respectively.
\subsection{Synoptic Optical Long-term Investigations of the Sun(SOLIS) - Vector-SpectroMagnetograph(VSM}
The Vector Spectromagnetograph \citep[VSM; see][]{Jones02} is part of the Synoptic Optical Long-term 
Investigations of the Sun (SOLIS) synoptic facility \citep[SOLIS; see][]{Keller03}. VSM is a full disk 
Stokes polarimeter. As part of daily synoptic observations, it takes four different observations in three 
spectral lines: Stokes $I$(intensity), $V$ (circular  polarization), $Q$, and $U$ (linear polarization) 
in photospheric spectral lines Fe {\Rmnum{1}} 630.15 nm and Fe {\Rmnum{1}} 630.25 nm , Stokes $I$ and $V$ 
in Fe {\Rmnum{1}} 630.15 nm and Fe {\Rmnum{1}} 630.25 nm, similar observations in chromospheric spectral 
line Ca {\Rmnum{2}} 854.2 nm, and Stokes $I$ in the He {\Rmnum{1}} 1083.0 nm line and the near by Si {\Rmnum{1}} 
spectral line. Observations of $I$, $Q$, $U$, and $V$ are used to construct full disk vector magnetograms, 
while $I-V$ observations are employed to create separate full disk longitudinal magnetograms in the photosphere 
and the chromosphere. The vector data are provided with a plate scale of one arc-second. The lower limits
for the noise levels are a few Gauss in the longitudinal and 70G in the transverse field measurements.

Quick-look (QL) vector magnetograms were created based on an algorithm by \citet{Auer:1977}. Beginning January 2012, 
QL vector magnetograms are created using weak-field approximation \citep{Ronan:1987}. The algorithm 
uses the Milne-Eddington model of solar atmosphere, which assumes that the magnetic field is uniform (no gradients) 
through the layer of spectral line formation \citep{Unno:1956}. It also assumes symmetric line profiles, disregards 
magneto-optical effects (e.g., Faraday rotation), and does not distinguish the contributions of magnetic and non-magnetic 
components in spectral line profiles (i.e., magnetic filling factor is set to unity). A complete inversion of the 
spectral data is performed later using a technique developed by \citet{Skumanich:1987}. This latter inversion 
(called ME magnetogram) also employs Milne-Eddington model of atmosphere, but solves for magneto-optical effects and
determines the magnetic filling factor i.e., (the fractional contribution of magnetic and non-magnetic components 
to each pixel). The ME inversion is only performed for pixels with spectral line profiles above the noise level. 
For pixels below the polarimetric noise threshold, the magnetic field parameters are set to zero.

From the measurements, the azimuths of transverse magnetic field can be determined with 180-degree
ambiguity. This ambiguity is resolved using the non-potential field calculation \citep[NPFC; see][]{Georgoulis05}.
The NPFC method was selected on the basis of a comparative investigation of several methods for 180-degree
ambiguity resolution \citep{Metcalf:2006}. Both QL and ME magnetograms can be used for potential and/or
force-free field extrapolation. However, in strong fields inside sunspots, the QL field strengths may
exhibit an erroneous decrease inside the sunspot umbra due to so-called magnetic saturation. For this study, we choose
to use fully inverted ME magnetograms. 
\section{Method}
{\refA{Photospheric field measurements are often subject to measurement errors. In addition to this, there are finite non-magnetic 
forces which make the data inconsistent as a boundary for a force-free field in the corona.}} In order to deal with these uncertainties, one has to: 1.) 
preprocess the surface measurements in order to make them compatible with a force-free field and 2.) keep a balance 
between the force-free constraint and deviation from the photospheric field measurements. Both methods contain 
free parameters, which have to be optimized for use with data from SOLIS/VSM and SDO/HMI. 
\subsection{Preprocessing of HMI and VSM data}
{\refA{To serve as suitable lower boundary condition for a force-free modeling, vector magnetograms have to be approximately 
flux balanced and on average a net tangential force acting on the boundary and shear stresses along axes lying on the boundary 
have to reduce to zero.}} We use dimensionless parameters, $\epsilon_{flux}$, $\epsilon_{force}$ and $\epsilon_{torque}$, 
to quantify such properties\citep{Wiegelmann06sak,tilaye09,Aly89,Molodenskii69}. 
Even if we choose a sufficiently flux balanced region ($\epsilon_{flux}$), we find that the force-free conditions 
$\epsilon_{force}\ll1$ and $\epsilon_{torque}\ll1$ are not usually fulfilled for measured vector magnetograms. In order 
to fulfill those conditions, we use preprocessing method as implemented in \citet{Wiegelmann06sak}. The preprocessing 
scheme of \citet{tilaye09} involves minimizing a two-dimensional functional of quadratic form in spherical geometry 
similar to
 \begin{displaymath} \vec{B}=\emph{argmin}(L_{p}),
\end{displaymath}
\begin{equation}
L_{p}=\mu_{1}L_{1}+\mu_{2}L_{2}+\mu_{3}L_{3}+\mu_{4}L_{4},\label{3}
\end{equation}
where $\vec{B}$ is the preprocessed surface magnetic field from the input observed field $\vec{B}_{obs}$. Each of the
constraints $L_{n}$ is weighted by an as yet undetermined factor $\mu_{n}$. The first term $(n=1)$ corresponds
to the force-balance condition, the next $(n=2)$ to the torque-free condition, and the last term $(n=4)$ controls
the smoothing. The explicit form of $L_{1}$, $L_{2}$, and $L_{4}$ can be found in \citet{tilaye09}. The term
$(n=3)$ controls the difference between measured and preprocessed vector fields.. 
\subsection{Optimization principle}
We solve the force-free equations (\ref{one}) and (\ref{two}) by optimization principle, as proposed
by \citet{Wheatland00} and generalized by \citet{Wiegelmann04} for cartesian geometry. The method minimizes
a joint measure of the normalized Lorentz forces and the divergence of the field throughout the volume of
interest, $V$. Throughout this minimization, the photospheric boundary of the model field $\vec{B}$ is matched
exactly to the observed $\vec{B}_{obs}$ and possibly preprocessed magnetogram values $\vec{B}$. Here, we use the optimization
approach for functional $(L_\mathrm{\omega})$ in spherical geometry \citep{Wiegelmann07,tilaye09} along with
the new method, which instead of an exact match enforces a minimal deviation between the photospheric boundary
of the model field $\vec{B}$ and the magnetogram field $\vec{B}_{obs}$ by adding an appropriate surface integral term
$L_{photo}$ \citep{Wiegelmann10,Tilaye:2010}. These terms are given by
\begin{displaymath} \vec{B}=\emph{argmin}(L_{\omega})
\end{displaymath}
\begin{equation}L_{\omega}=L_{f}+L_{d}+\nu L_{photo} \label{4}
\end{equation}
\begin{displaymath} L_{f}=\int_{V}\omega_{f}(r,\theta,\phi)B^{-2}\big|(\nabla\times {\vec{B}})\times
{\vec{B}}\big|^2  r^2\sin\theta dr d\theta d\phi
\end{displaymath}
\begin{displaymath}L_{d}=\int_{V}\omega_{d}(r,\theta,\phi)\big|\nabla\cdot {\vec{B}}\big|^2
  r^2\sin\theta dr d\theta d\phi
\end{displaymath}
\begin{displaymath}L_{photo}=\int_{S}\big(\vec{B}-\vec{B}_{obs}\big)\cdot\vec{W}(\theta,\phi)\cdot\big(
\vec{B}-\vec{B}_{obs}\big) r^{2}\sin\theta d\theta d\phi
\end{displaymath}
where $L_{f}$ and $L_{d}$ measure how well the force-free Eqs.~(\ref{one}) and divergence-free (\ref{two}) conditions
are fulfilled, respectively, and both $\omega_{f}(r,\theta,\phi)$ and $\omega_{d}(r,\theta,\phi)$ are weighting
functions. The weighting functions $\omega_{f}$ and $\omega_{d}$ in $L_{f}$ and $L_{d}$ in Eq.~(\ref{4}) are chosen to 
be unity within the inner physical domain $V'$ and decline with a cosine profile in the buffer boundary region 
\citep{Wiegelmann04,tilaye09,Tilaye:2012a}. They reach a zero value at the boundary of the outer volume $V$. The distance 
between the boundaries of $V'$ and  $V$ is chosen to be $nd=10$ grid points wide. The third integral, $L_{photo}$, is 
the surface integral over the photosphere which allows us to relax the field on the photosphere towards force-free 
solution without too much deviation from the original surface field data. 

$\vec{W}(\theta,\phi)$ is a space-dependent diagonal matrix the element of which are inverse proportional to 
the estimated squared measurement error of the respective field component. In principle one could compute $\vec{W}$ 
from the measurement noise and errors obtained from the inversion of measured Stokes profiles to field components. 
Until these quantities become available, a reasonable assumption is that the magnetic field is measured in strong field 
regions more accurately than in the weak field and that the error in the photospheric transverse field is at least one 
order of magnitude higher as the line-of-sight component. Appropriate choices to optimize $\nu$ and $\vec{W}$ for use with 
SDO/HMI\citep{Wiegelmann:2012} and SOLIS/VSM\citep{Tilaye:2010} magnetograms have been investigated. For a 
detailed description of the current code implementation, we refer to \citet{Wiegelmann10} and \citet{Tilaye:2010}.
\section{Results}
Within this work, we use the full disk data from SOLIS/VSM and SDO/HMI instruments obsrved on November 09 2011 around 17:45UT. 
During this observation there were four active regions (ARs 11338, 11339, 11341 and 11342) along with other 
smaller sunspots spreading on the disk. {\refA{To accommodate the connectivity between those ARs and their surroundings, 
we adopt a non uniform spherical grid $r$, $\theta$, $\phi$ with $n_{r}=225$, $n_{\theta}=375$, $n_{\phi}=425$ grid points in the direction 
of radius, latitude, and longitude, respectively, with the field of view of 
$[r_{\text{min}}=1R_{\sun}:r_{\text{max}}=2R_{\sun}]\times[\theta_{\text{min}}=-50^{\circ}:\theta_{\text{max}}=50^{\circ}]\times[\phi_{\text{min}}=90^{\circ}:\phi_{\text{max}}=270^{\circ}]$.}} 
Given the twice as large plate scale of VSM, we bin the HMI vector maps to the resolution of VSM in order to compare the photospheric magnetic 
field and subsequent force free modeling.
\begin{figure*}[htp!]
   \centering
 \includegraphics[viewport=10 5 828 800,clip,height=13cm,width=15.0cm]{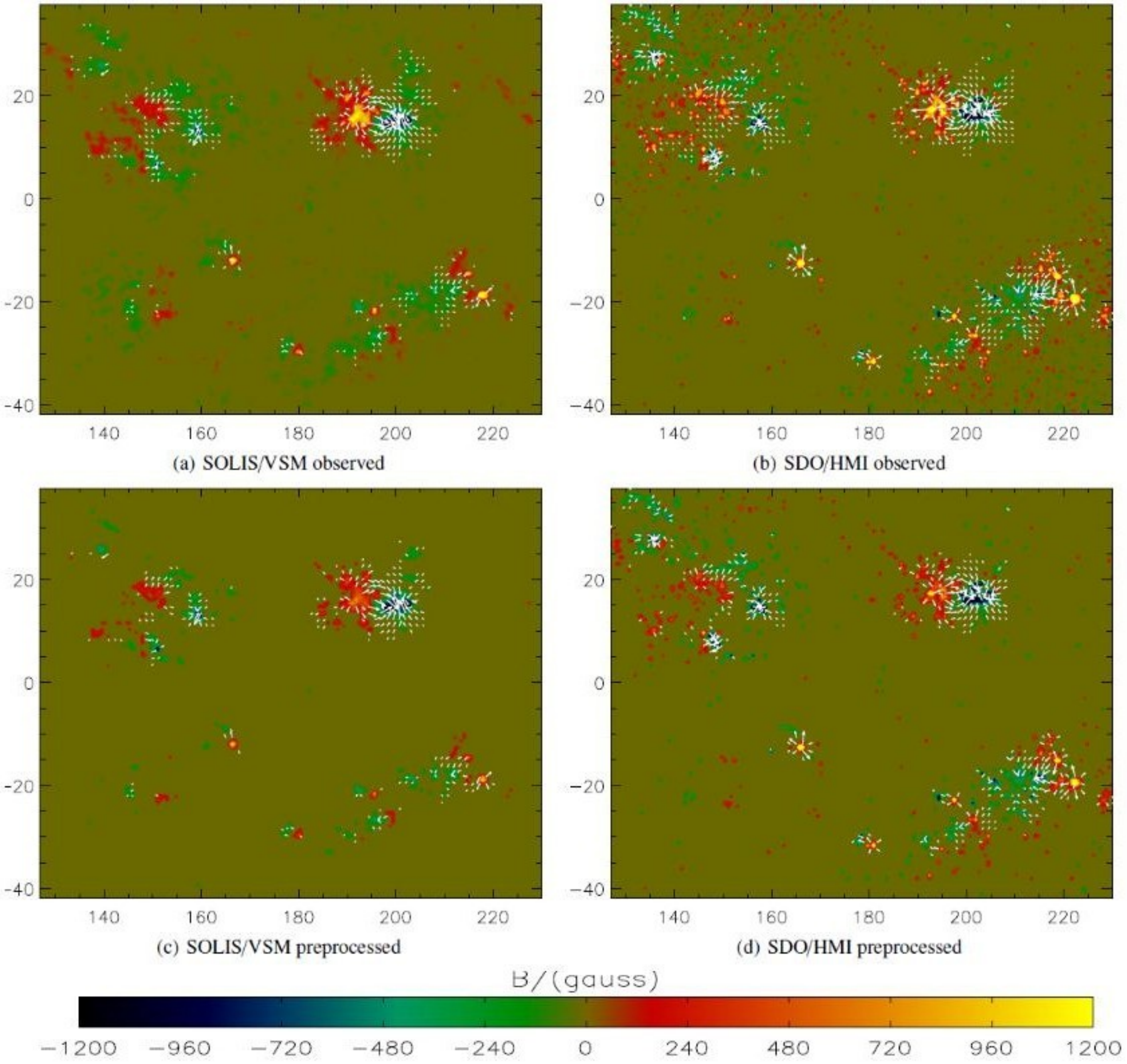}
\caption{Magnetic vector maps of VSM and HMI on part of the lower boundary. The color coding shows $B_{r}$ on the photosphere 
and the white arrow indicates the transverse components of the field. The vertical and horizontal axes show latitude, $\theta$
and longitude, $\phi$ on the photosphere respectively. }\label{fig1}
 \end{figure*}
To deal with vector magnetogram data being inconsistent with the force-free assumption, we use a preprocessing routine 
in spherical geometry, which derives suitable boundary conditions for force-free modeling from the measured photospheric 
data. Applying this procedure to both SDO/HMI and SOLIS/VSM reduces $\epsilon_{force}$ and $\epsilon_{torque}$ further 
significantly. The two quantities are very well below unity after preprocessing, which gives us some confidence that the 
data might serve as suitable boundary condition for a force-free modeling. Doing this, we do not intend to suppress the 
existing forces in the photosphere. Instead, we try to approximate the magnetic field at a chromospheric level where 
magnetic forces are expected to be much smaller than in the layers below. Both vector magnetograms are almost flux balanced 
and the field of view was large enough to cover the full-disk. The unsigned magnetic flux of longitudinal surface magnetic 
field from HMI is 1.57 times that of VSM magnetogram. This is in agreement with recent comparative study by \citet{Pietarila:2012}, 
who found that the factor to convert SOLIS/VSM to SDO/HMI increases with flux density from  about 1 (weak fields) to about 1.5 
(strong fields) {\refA{for the line-of-sight full disk magnetograms}}. HMI inverts weak field regions, however, for VSM zeros are assigned 
to pixels where the measured polarization signal is too weak to perform a reliable inversion. Disregarding these “zero-pixels”, 
about 20\% of the total number of pixels in the HMI and VSM full disk vector maps are remaining for comparison. HMI is found to 
detect most transverse field.

We used a standard preprocessing parameter set $\mu_{1} = \mu_{2} = 1$ and $\mu_{3}=0.001$, which are similar to the values 
calculated from vector data used in previous studies\citep{Wiegelmann:2012} for HMI data in Cartesian coordinates. 
Table~\ref{table1} lists the values of dimensionless parameters for the used HMI and VSM data-sets. In this study, we have 
found that the optimal value for smoothing parameter is $\mu_{4}=0.05$ for full-disk HMI data. These parameters control the 
amount of force-freeness, torque-freeness, nearness to the actually observed data and smoothing, respectively. As the result 
of parameter study, \citet{Tilaye:2010} have found $\mu_{1} = \mu_{2} = 1$, $\mu_{3}=0.03$ and $\mu_{4}=0.45$ as optimal for 
full-disk VSM data. 

The preprocessing influences the structure of the magnetic vector data. It does not only smooths $\vec{B}_{t}$ (transverse field) 
but also alters its values in order to reduce the net force and torque. The change in $\vec{B}_{t}$ is more pronounced than 
the radial component $\vec{B}_{r}$ (radial field) since $\vec{B}_{t}$ is measured with lower accuracy than the longitudinal magnetic 
field. Figure~\ref{fig1} shows the preprocessed and observed surface vector magnetic field obtained from SDO/HMI and SOLIS/VSM 
magnetograms. To identify the similarity of vector components from HMI and VSM on the bottom surface, we calculate their pixel-wise 
correlations before and after preprocessing. The correlation were calculated from
\begin{equation}
C_\mathrm{ vec}= \frac{ \sum_i \vec{v}_{i} \cdot \vec{u}_{i}}{ \Big( \sum_i |\vec{v}_{i}|^2 \sum_i
|\vec{u}_{i}|^2 \Big)^{1/2}}, \label{6}
\end{equation}
where $\vec{v}_{i}$ and $\vec{u}_{i}$ are the vectors at each grid point $i$ on the bottom surface. If the vector fields are 
identical, then $C_{vec}=1$; if $\vec{v}_{i}\perp \vec{u}_{i}$ , then $C_{vec}=0$. Table~\ref{table2} shows the correlation 
($C_{vec}$) of the 2D surface magnetic field vectors of observed and preprocessed data from HMI and VSM for the radial and 
transverse components. The vector correlation between $\vec{B}_{t}$ in the preprocessed HMI and VSM surface vector maps is 
clearly more closer to unity than the corresponding surface vector maps without preprocessing. There is no such difference 
in correlations between $\vec{B}_{r}$ before and after preprocessing. This is to be expected since the preprocessing scheme 
only smooths the longitudinal field while it smooths and alters the transverse field. The mean value of the changes due to 
preprocessing in the longitudinal field is $10^{-3}$G and for the transverse field on the order of 10 G, i. e., well within 
the measurement uncertainty of the HMI and VSM.
\begin{figure}
   \centering
\includegraphics[viewport=12 20 480 795,clip,height=12.2cm,width=8.9cm]{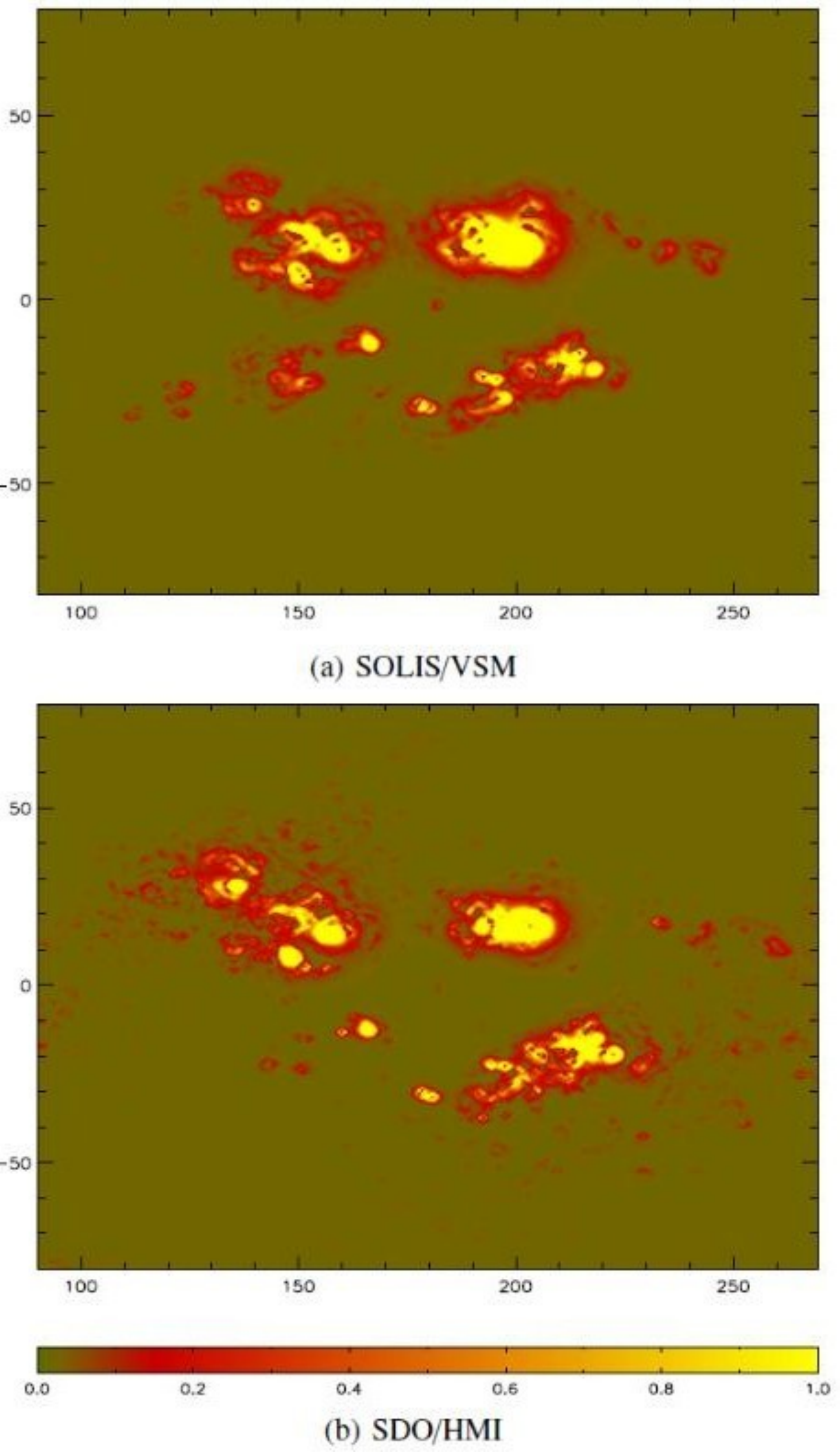}
\caption{Mask function for magnetic vector field distribution on full disk from (a) VSM and (b)  HMI. The vertical 
and horizontal axes show latitude, $\theta$ and longitude, $\phi$ on the photosphere respectively.}\label{fig2}
 \end{figure}
\begin{center}
\begin{table}
\caption{Flux-balance, force and torque free parameters of SOLIS/VSM and SDO/MHI full disk magnetograms.}
\label{table1}
\begin{tabular}{ccccc}
 \hline \hline
 \multilineL{Data set}&\multilineL{$\epsilon_{flux}$} &\multilineL{$\epsilon_{force}$}& \multilineL{$\epsilon_{torque}$} \\
\hline
HMI observed &$-0.0621$ &$0.1305$ & $0.1773$\\
HMI preprocessed &$-0.0313$&$0.0001$&$0.0002$\\
SOLIS observed &$-0.0857$&$0.4571$&$0.2947$\\
SOLIS preprocessed &$-0.0460$&$0.0015$&$0.0007$\\
\hline
\end{tabular}
\end{table}
\end{center}  
Before we perform nonlinear force-free extrapolations we use the preprocessed radial component $\vec{B}_{r}$ of the VSM and 
HMI-data to compute the corresponding potential fields using spherical harmonic expansion for initializing our code. We 
implement the new term $L_{photo}$ in Eq.~(\ref{4}) to work with boundary data of different noise levels and qualities or even 
neglect some data points completely. SOLIS/VSM provides full-disk vector-magnetograms, but for some individual pixels the inversion 
from line profiles to field values may not have been successfully inverted and field data there will be missing for these pixels. 
Since the old code without the $L_{photo}$ term requires complete boundary information, it cannot be applied to this set 
of SOLIS/VSM data. In our new code, these data gaps are treated by setting $W=0$ for these pixels in Eq.~(\ref{4})
\citep{Wiegelmann10,Tilaye:2010}. For those pixels, for which $\vec{B}_{obs}$ was successfully inverted, we allow 
deviations between the model field $\vec{B}$ and the input fields observed $\vec{B}_{obs}$ surface field using 
Eq.~(\ref{4}), so that the model field can be iterated closer to a force-free solution even if the observations are 
inconsistent. 

The improved optimization scheme allows us to relax the magnetic field also on the lower boundary. The relaxation of 
the lower boundary introduces a further modification of the vector data, in addition to that by the preprocessing applied 
before. The mean modification of the longitudinal field due to the relaxation of the lower boundary is $10^{-4}$G and absolute 
values are on the order of 1 G. The mean changes of the transverse field are on the order of 10G and absolute values can be 
several 100 G. Given the noise levels of HMI and VSM measurements of the longitudinal ($\approx10$G and a few G, respectively) 
and transverse field ($\approx100$G and $\gtrsim70$G, respectively), the modifications are on the order of the measurement error. 

For nonlinear force-free fields we minimize the functional Eq.~(\ref{4}). In order to control the speed with which the lower 
boundary is injected during the extrapolation, we vary the Langrangian multiplier $\nu$. Unless an exact error computation 
becomes available from inversion and ambiguity removal of the photospheric magnetic field vector, a reasonable assumption 
is that the field is measured more accurately in strong field regions and one can carry out computations with the mask $\propto B_{t}$ 
and $\propto B^{2}_{t}$. We choose a mask function of $W=\big(B_{t}/max(B_{t})\big)^{2}$, which gives more weight to strong 
field regions than to weak ones as investigated  in \citet{Wiegelmann:2012}. Figure~\ref{fig2} shows surface distribution 
of mask function $W$ for VSM and HMI full-disk data. For strong field regions  $W$ is close to unity and decline to zero in 
weaker field regions. We vary the Langrangian multiplier $\nu$ between $0.1$ and $0.0001$ to investigate the optimal parameter 
for HMI full-disk data. To evaluate how well the force-free and divergence-free condition are satisfied for different Langrangian 
multiplier $\nu$, we monitor a number of expressions, such as $L_{f}$, $L_{d}$ and 
\begin{equation}
\sigma_{j}=\Big(\sum_{i}\frac{|\vec{J}_{i}\times \vec{B}_{i}|}{B_{i}}\Big)/\sum_{i}J_{i}, \label{7}
\end{equation}
where $\sigma_{j}$ is the sine of the current weighted average angle between the magnetic field $\vec{B}$ and electric 
current $\vec{J}$. 
\begin{table}
\caption{The correlations between the components of surface fields from HMI and VSM data.}
\label{table2}
\begin{tabular}{cccc}
  \hline \hline
 & v & u & $C_\mathrm{vec}$\\
\hline
No preprocessing &$(\vec{B}_{HMI})_{r}$&$ (\vec{B}_{VSM})_{r}$ &$0.947$\\
No preprocessing & $(\vec{B}_{HMI})_{t}$&$ (\vec{B}_{VSM})_{t}$ &$0.893$\\
Preprocessed & $(\vec{B}_{HMI})_{r}$&$ (\vec{B}_{VSM})_{r}$ &$0.965$\\
Preprocessed & $(\vec{B}_{HMI})_{t}$&$ (\vec{B}_{VSM})_{t}$ &$0.951$\\
\hline
\end{tabular}
\end{table}
\begin{table}
\caption{Evaluation of force-free field models from preprocessed HMI data. The first column names the model and in 
column 2 shows the used Langrangian multipliers. Columns 3-5 show different force-free consistency evaluations. Column 6 
shows the ratio of NLFFF energy density to the corresponding potential energy density and column 7 the computing time.}
\label{table3}
\begin{tabular}{cccccc}
  \hline \hline
 $\nu$ & $L_{f}$ & $L_{d}$ & $sin^{-1}(\sigma_{i})$& $E/E_{\text{pot}}$&Time\\
\hline
$0.1$&$21.7$ &$13.4$&$25.8^{\circ}$&$1.06$&2h:17\,min\\
$0.05$&$19.8$ &$10.7$&$18.1^{\circ}$&$1.12$&3h:31\,min\\
$0.001$&$2.9$ &$1.5$&$4.8^{\circ}$&$1.22$&4h:39\,min\\
$0.005$&$5.2$ &$3.9$&$8.9^{\circ}$&$1.23$&11h:47\,min\\
$0.0001$&$7.7$ &$4.3$&$10.2^{\circ}$&$1.26$&48h:53\,min\\
\hline
\end{tabular}
\end{table}
\begin{figure*}[htp!]
   \centering
 \includegraphics[bb=10 10 870 830,clip,height=14.2cm,width=14.2cm]{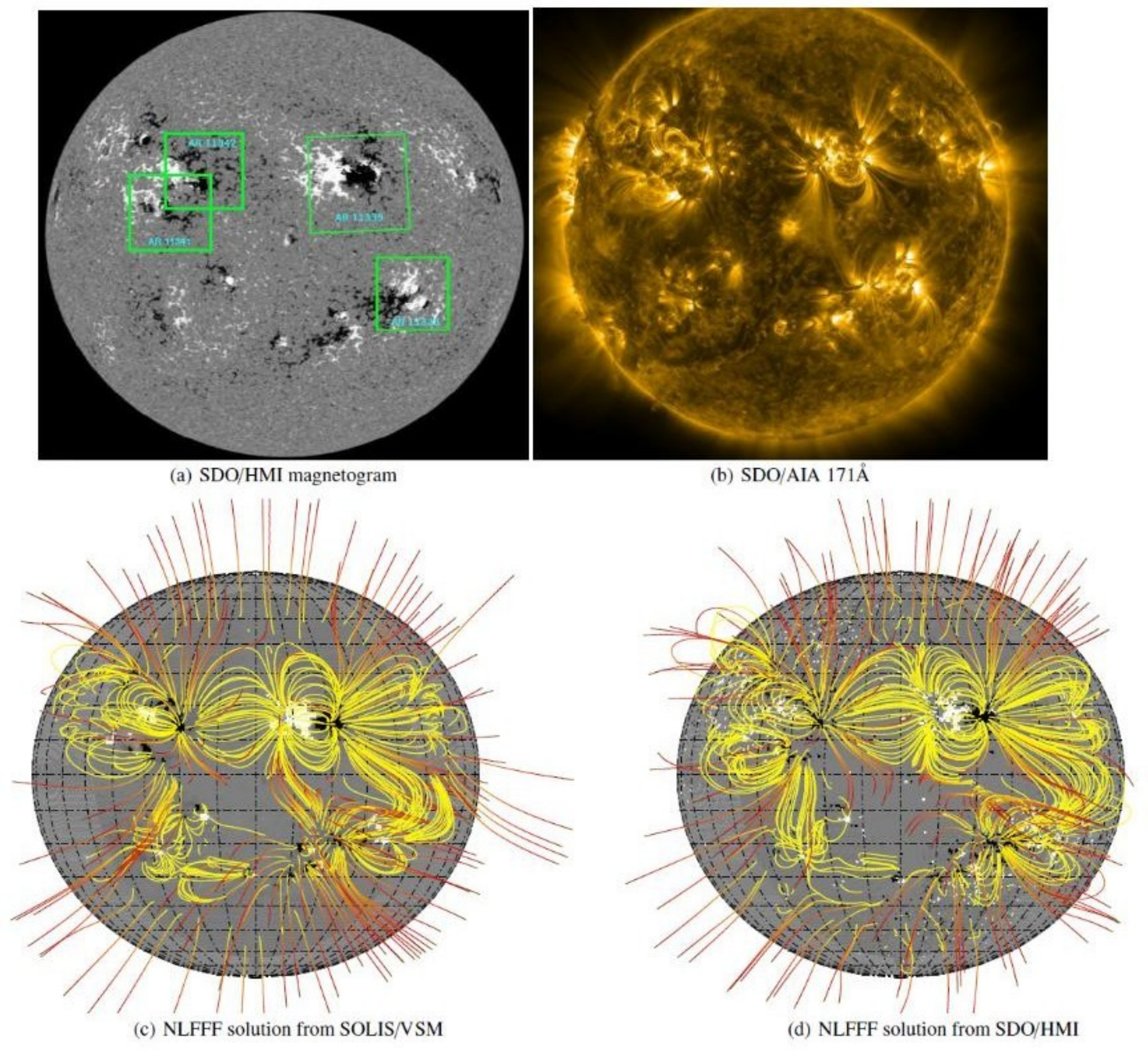}
\caption{{\refA{a) SDO/HMI and b) AIA images and their respective selected magnetic field lines plots reconstructed from c) SOLIS and d) HMI 
magnetograms using nonlinear force-free modelling. The color coding shows $B_{r}$ on the photosphere. Yellow field lines represent 
closed field lines, while field lines changing in color from yellow to brown (from bottom to the top) represent the open ones. 
The gray area indicates the region where magnetic field values are close to zero.}}}\label{fig3}
 \end{figure*}

For a sufficient small Lagrangian multiplier $\nu = 0.001$ we found that the resulting coronal fields are force and
divergence free compared to other values as shown in Table~\ref{table3}. The weighted angle between the magnetic field 
and electric current is about $5^{\circ}$ for $\nu = 0.001$. Injecting the boundary faster by choosing a higher Lagrangian 
multiplier ($\nu = 0.1$) speeds up the computation, but the residual forces are higher and current and field are not 
well aligned as investigated by \citet{Wiegelmann:2012} for single AR. 

To understand the physics of solar flares, including the local reorganization of the magnetic field and the
acceleration of energetic particles, one has to estimate the free magnetic energy  available for these
phenomena {\refA{\citep{Regnier:2007a,Aschwanden:2008,Schrijver:2009a}.}} This is the free energy that can be converted into 
kinetic and thermal energy. From the energy budget and the observed magnetic activity in the active region, \citet{Regnier} 
and \citet{Thalmann} investigated the free energy above the minimum-energy state for the flare process. We estimate the 
free magnetic energy  to be the difference between the extrapolated force-free fields and the potential field with the 
same normal boundary conditions in the photosphere. We therefore estimate the upper limit to the free magnetic energy  
associated with coronal currents of the form
\begin{equation}
E_\mathrm{free}=\frac{1}{8\pi}\int_{V}\Big(B_{nlff}^{2}-B_{pot}^{2}\Big)r^{2}sin\theta dr d\theta d\phi, \label{ten}
\end{equation}
\begin{center}
\begin{table}
\caption{The magnetic energy  associated with extrapolated NLFFF field configurations from full disk SDO/HMI and SOLIS/VSM data.}
\label{table4}
\begin{tabular}{ccc}
 \hline \hline
Model & $E_{nlff}(10^{33}erg)$& $E_\mathrm{free}(10^{33}erg)$\\
\hline
SOLIS/VSM &$8.609$&$1.375$\\
SDO/HMI &$8.913$&$1.607$\\
\hline
\end{tabular}
\end{table}
\end{center}
where $B_{pot}$ and $B_{nlff}$ represent the potential and NLFFF magnetic field,respectively. The magnetic energy 
densities associated with the potential field configurations from SDO/HMI and SOLIS/VSM data are found to be 
$7.306\times10^{33}erg$ and $7.234\times10^{33}erg$, respectively. This has to be expected as the unsigned magnetic 
flux of longitudinal surface magnetic field from HMI is greater than that of VSM magnetogram. The magnetic energy  
of NLFFF obtained from HMI data is greater that the one obtained from VSM data as shown in Table~\ref{table4}. This is due 
to the fact that HMI data has more longitudinal unsigned  magnetic flux and detects more transverse field than VSM. 
To study the influence of the use of preprocessed boundary data from the observed boundary fields on the estimation 
of free-magnetic energy , we have computed the magnetic energy  associated with the potential field and NLFFF configurations 
from the original SDO/HMI data without preprocessing and with preprocessing. The case for SOLIS/VSM has been studied 
by \citet{Tilaye:2012}. As preprocessing procedure filters out small scale surface field fluctuations, the 
magnetic energy  associated with NLFFF obtained from preprocessed SDO/HMI boundary data is smaller than the 
one without preprocessing. Obviously, the potential field energies of boundary data with and without preprocessing 
are close in value, since the potential field calculation makes use of the radial magnetic field component which 
is not affected too much by preprocessing procedure. The computed magnetic energy  from SDO/HMI original data without 
preprocessing is about $9.067\times10^{33}\textrm{erg}$, which is about $1.7\%$ higher than the one obtained from 
preprocessed and modified observational HMI boundary data. However, this energy does not correspond to the nonlinear
force-free magnetic field solution since the original boundary data without preprocessing is not a consistent boundary 
condition for NLFFF modeling\citep{Tilaye:2012}. 

We investigate the magnetic field configurations of the VSM and HMI models by comparing the vector components. We
calculate the vector correlation (using Equation ~(\ref{6}) in the computational volume) of the potential fields and 
the NLFFF fields. The average vector correlation between the potential fields based on the HMI and VSM data is 0.97. 
The average vector correlation between the NLFFF fields of HMI and VSM data is 0.94. Figure~\ref{fig3} a. and b. show 
the surface radial magnetic field component observed by HMI instrument on November 09 2011 and the corresponding 
AIA (Atmospheric Imaging Assembly) image in 171\AA{}, respectively. {\refA{The magnetic field lines obtained from nonlinear 
force-free extrapolation based on HMI and VSM data have good correlations as shown in Figure~\ref{fig3} c. and d., with the foot points of 
the field lines from the two magnetograms are identical.}} However, there are some differences. For example, extrapolated field lines from SDO/HMI 
magnetogram (Figure 3d) do not show transequatorial loops connecting trailing polarity of NOAA AR 11339 (west of central meridian in northern hemisphere) 
and trailing polarity of AR11338 (southern hemisphere). This transequatorial loop is well represented by NLFFF extrapolation 
based on SOLIS/VSM. This difference can be attributed to presence of a patch of weak fields between two active regions (in  SDO/HMI data). 
With this weak field patch, SDO/HMI model tends to close field lines originating in trailing polarity of AR11339, while SOLIS/VSM model 
extends them to AR11338. Both extrapolations indicate loops connecting AR11339 and AR11342 (East of central meridian in Northern hemisphere). 
Although SDO/AIA image (Figure 3b) does not show coronal loops connecting these two active region, such loops are clearly visible 
in images taken by X-ray Telescope on Hinode. These loops appear to fit better field lines from SOLIS/VSM model. Despite a relatively 
good visual agreement in extrapolated fields, the models show some notable disagreement in derived magnetic energy. Thus, for example, 
the estimated free magnetic energy  obtained from SDO/HMI is $14.4\%$ higher than that of SOLIS/VSM. This 
is due to the fact that HMI data includes small scale magnetic fields measurements. 
\begin{figure}[htp!]
   \centering
\includegraphics[viewport=5 10 500 795,clip,height=13.2cm,width=8.5cm]{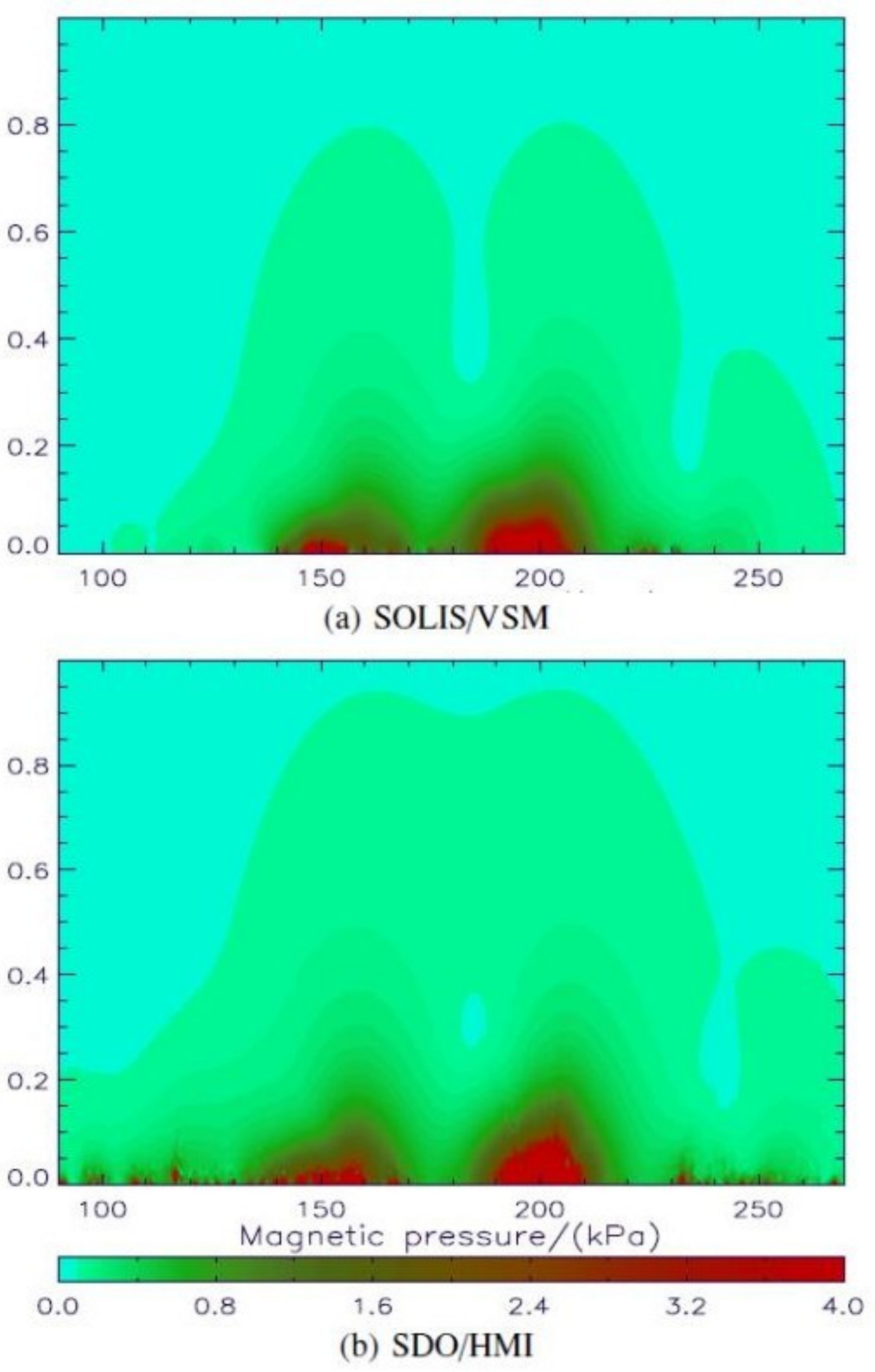}
\caption{Magnetic pressure $p_{m}$ in the longitudinal cross-section at $\theta=20^{\circ}$ for the (a) VSM and 
(b)  HMI. The vertical and horizontal axes show radial distance in solar radius and longitude, $\phi$ on the 
photosphere, respectively.}\label{fig4}
 \end{figure}
\begin{figure}
   \centering
\includegraphics[viewport=5 10 480 795,clip,height=12.5cm,width=9.5cm]{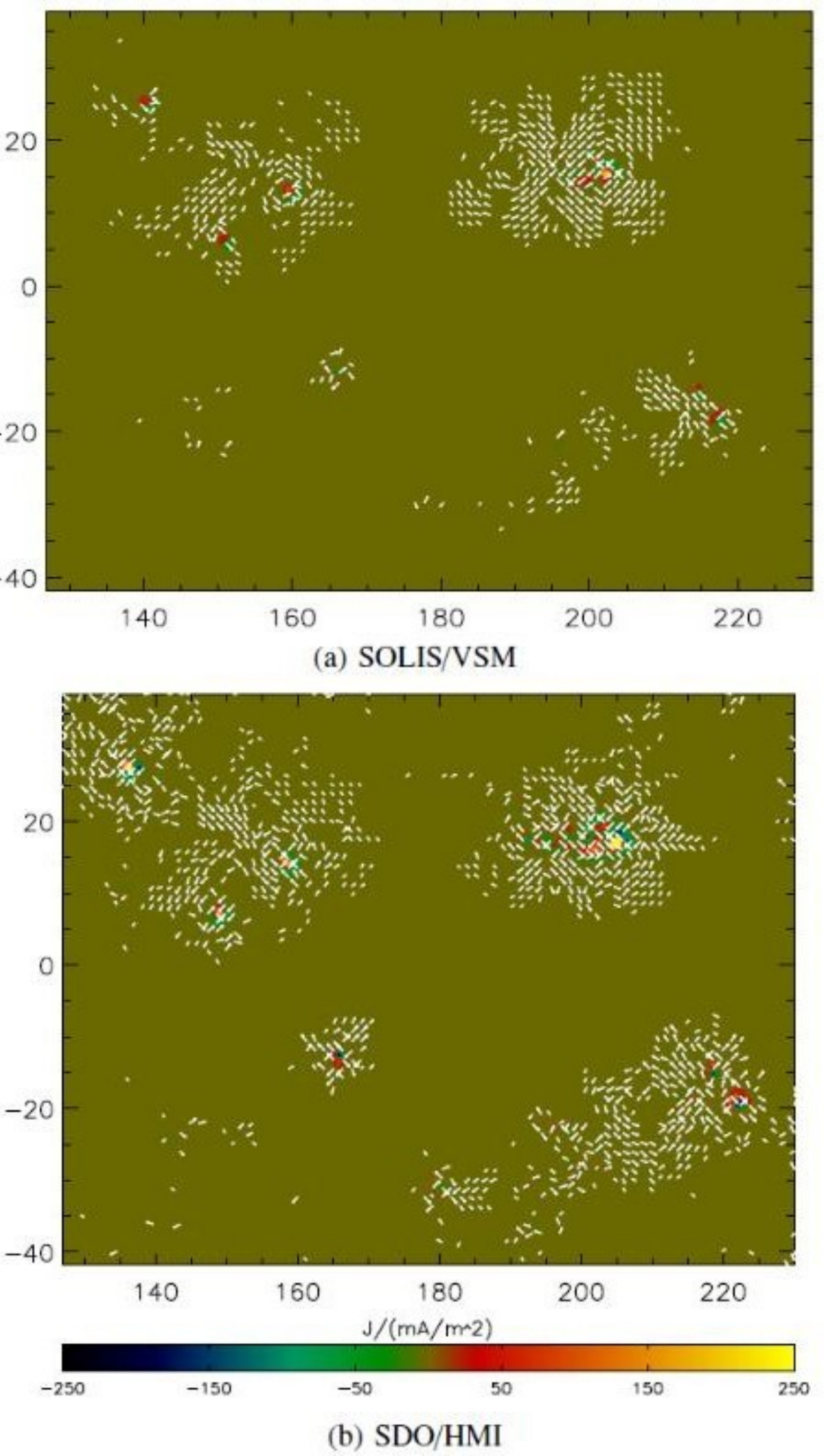}
\caption{Vector plot of the radial component of electric current density and vector field plot of transverse 
component of electric current density with white arrows. The colour coding shows $J_{r}$ on the photosphere. 
The vertical and horizontal axes show latitude, $\theta$ and longitude, $\phi$ on the photosphere respectively.}\label{fig5}
 \end{figure}
We study the magnetic pressure, $p_{m}$, in a longitudinal cross-section at about $\theta=20^{\circ}$ as shown in 
Figure~\ref{fig4}. The overall pattern of $p_{m}$ appears to be the same when calculated from the HMI and VSM NLFFF model volume. 
In this study we found that the magnetic pressure of NLFFF model field from HMI is greater than that of VSM for same locations
in the cross section. {\refA{This is expected, since the magnetic pressure is proportional to the magnetic energy density as the magnetic energy 
of NLFFF model field from HMI is larger than that of VSM.}}  The surface radial ($J_{r}$) and transverse ($\vec{J}_{t}$) electric 
current densities of the NLFFF field models based on HMI and VSM data are shown in in Figure~\ref{fig5}. {\refA{The value of the total radial 
surface  electric current density flux of the NLFFF field models based on HMI is greater that than that of VSM. It agrees with fact that the HMI 
instrument measures more transverse magnetic field than that of VSM instrument.}} The transverse surface electric current density of 
the NLFFF field model based on HMI spreads more around the active regions than that of VSM as shown in Figure~\ref{fig5}. This could 
reflect the fact \citep[see]{Pietarila:2012} that the scaling factor between SOLIS/VSM and SDO/HMI is different for weak and strong 
fluxes. This difference is scaling factor may act as a weighting function when comparing electric currents derived from two models. 
In addition, the vector correlations of the radial and transverse surface electric current densities of the NLFFF field models based 
on HMI and VSM are 0.96 and 0.88, respectively. This indicate that there is more pronounced discrepancy in transverse electric current 
densities than radial one.
\section{Conclusion and outlook}
We have investigated the coronal magnetic field associated with full solar disk on November 09 2011 by analysing 
SDO/HMI and SOLIS/VSM data. We carried out nonlinear force-free coronal field extrapolations of a full disk magnetograms. 
The vector magnetogram is almost perfectly flux balanced and the field of view was large enough to cover all the weak field 
surrounding the active regions. Both conditions are necessary in order to carry out meaningful force-free computations. We 
have used the optimization method for the reconstruction of nonlinear force-free coronal magnetic fields in spherical 
geometry \citep{Wiegelmann07,tilaye09} to campare the final NLFFF model field solution from HMI and VSM full disk data. 

We have found that the optimal value for smoothing parameter is $\mu_{4}=0.05$ for full-disk HMI data for the purpose of 
preprocessing. We conclude that the choice $\nu=0.001$ is the optimal choices for HMI full disk data set for our new code 
as investigated in \citet{Wiegelmann:2012}.

The magnetic field lines obtained from nonlinear force-free extrapolation based on HMI and VSM data have good correlations. However, 
the models show some disagreement on the estimated relative free magnetic energy which can be released during explosive events and 
surface electric current density. {\refA{ Reconstructed magnetic field based on SDO/HMI data have more contents of total magnetic energy, 
free magnetic energy, the longitudinal distribution of the magnetic pressure and surface electric current density compared to SOLIS/VSM data.}} 
Since the disagreement in free energy can be attributed to presence of weaker transverse fields in SDO/HMI measurements, it is not clear how 
important is the found (14.4\%) difference in {\refA{free magnetic energy}} for flare and CME processes originating in magnetic fields higher 
in the corona. This aspect deserves a separate study. 
\begin{acknowledgements} The authors thank the anonymous referee for helpful comments. Data are courtesy of NASA/SDO and the AIA and 
HMI science teams. SOLIS/VSM vector magnetograms are produced cooperatively by NSF/NSO and NASA/LWS. The National Solar Observatory 
(NSO) is operated by the Association of Universities for Research in Astronomy, Inc., under cooperative agreement with the National Science Foundation. This 
work was supported by NASA grant NNX07AU64G and the work of T. Wiegelmann was supported by DLR-grant $50$ OC $453$  $0501$.
\end{acknowledgements}
\bibliographystyle{aa}
\bibliography{aa_bibtex}

\begin{thebibliography}{54}
\expandafter\ifx\csname natexlab\endcsname\relax\def\natexlab#1{#1}\fi

\bibitem[{{Aly}(1989)}]{Aly89}
{Aly}, J.~J. 1989, Sol. Phys., 120, 19

\bibitem[{{Amari} \& {Aly}(2010)}]{Amari:2010}
{Amari}, T. \& {Aly}, J. 2010, A\&A, 522, A52+

\bibitem[{{Amari} {et~al.}(1997){Amari}, {Aly}, {Luciani}, {Boulmezaoud}, \&
  {Mikic}}]{Amari97}
{Amari}, T., {Aly}, J.~J., {Luciani}, J.~F., {Boulmezaoud}, T.~Z., \& {Mikic},
  Z. 1997, Sol. Phys., 174, 129

\bibitem[{{Amari} {et~al.}(2006){Amari}, {Boulmezaoud}, \& {Aly}}]{Amari}
{Amari}, T., {Boulmezaoud}, T.~Z., \& {Aly}, J.~J. 2006, A\&A, 446, 691

\bibitem[{{Amari} {et~al.}(1999){Amari}, {Boulmezaoud}, \& {Mikic}}]{Amari99}
{Amari}, T., {Boulmezaoud}, T.~Z., \& {Mikic}, Z. 1999, A\&A, 350, 1051

\bibitem[{{Aschwanden}(2008)}]{Aschwanden:2008}
{Aschwanden}, M.~J. 2008, Journal of Astrophysics and Astronomy, 29, 115

\bibitem[{{Auer} {et~al.}(1977){Auer}, {House}, \& {Heasley}}]{Auer:1977}
{Auer}, L.~H., {House}, L.~L., \& {Heasley}, J.~N. 1977, Sol. Phys., 55, 47

\bibitem[{{Borrero} {et~al.}(2011){Borrero}, {Tomczyk}, {Kubo},
  {Socas-Navarro}, {Schou}, {Couvidat}, \& {Bogart}}]{Borrero:2011}
{Borrero}, J.~M., {Tomczyk}, S., {Kubo}, M., {et~al.} 2011, Sol. Phys., 273,
  267

\bibitem[{{Canou} \& {Amari}(2010)}]{Canou:2010}
{Canou}, A. \& {Amari}, T. 2010, ApJ, 715, 1566

\bibitem[{{Canou} {et~al.}(2009){Canou}, {Amari}, {Bommier}, {Schmieder},
  {Aulanier}, \& {Li}}]{Canou:2009}
{Canou}, A., {Amari}, T., {Bommier}, V., {et~al.} 2009, ApJL, 693, L27

\bibitem[{{DeRosa} {et~al.}(2009){DeRosa}, {Schrijver}, {Barnes}, {Leka},
  {Lites}, {Aschwanden}, {Amari}, {Canou}, {McTiernan}, {R{\'e}gnier},
  {Thalmann}, {Valori}, {Wheatland}, {Wiegelmann}, {Cheung}, {Conlon},
  {Fuhrmann}, {Inhester}, \& {Tadesse}}]{DeRosa}
{DeRosa}, M.~L., {Schrijver}, C.~J., {Barnes}, G., {et~al.} 2009, \apj, 696,
  1780

\bibitem[{{Gary}(2001)}]{Gary}
{Gary}, G.~A. 2001, Sol. Phys., 203, 71

\bibitem[{{Georgoulis}(2005)}]{Georgoulis05}
{Georgoulis}, M.~K. 2005, ApJL, 629, L69

\bibitem[{{Guo} {et~al.}(2010){Guo}, {Ding}, {Schmieder}, {Li},
  {T{\"o}r{\"o}k}, \& {Wiegelmann}}]{Guo:2010}
{Guo}, Y., {Ding}, M.~D., {Schmieder}, B., {et~al.} 2010, ApJL, 725, L38

\bibitem[{{Inhester} \& {Wiegelmann}(2006)}]{Inhester06}
{Inhester}, B. \& {Wiegelmann}, T. 2006, Sol. Phys., 235, 201

\bibitem[{{Jiang} \& {Feng}(2012)}]{Jiang:2012}
{Jiang}, C. \& {Feng}, X. 2012, ApJ, 749, 135

\bibitem[{{Jones} {et~al.}(2002){Jones}, {Harvey}, {Henney}, {Hill}, \&
  {Keller}}]{Jones02}
{Jones}, H.~P., {Harvey}, J.~W., {Henney}, C.~J., {Hill}, F., \& {Keller},
  U.~C. 2002, ESA SP, 505, 15

\bibitem[{{Keller} {et~al.}(2003){Keller}, {Harvey}, \& {Giampapa}}]{Keller03}
{Keller}, U.~C., {Harvey}, J.~W., \& {Giampapa}, M.~S. 2003, 4853, 194

\bibitem[{{Leka} {et~al.}(2009){Leka}, {Barnes}, {Crouch}, {Metcalf}, {Gary},
  {Jing}, \& {Liu}}]{Leka:2009}
{Leka}, K.~D., {Barnes}, G., {Crouch}, A.~D., {et~al.} 2009, Sol. Phys., 260,
  83

\bibitem[{{Metcalf}(1994)}]{Metcalf:1994}
{Metcalf}, T.~R. 1994, Sol. Phys., 155, 235

\bibitem[{{Metcalf} {et~al.}(2008){Metcalf}, {De Rosa}, {Schrijver}, {Barnes},
  {van Ballegooijen}, {Wiegelmann}, {Wheatland}, {Valori}, \&
  {McTtiernan}}]{Metcalf:2008}
{Metcalf}, T.~R., {De Rosa}, M.~L., {Schrijver}, C.~J., {et~al.} 2008, Sol.
  Phys., 247, 269

\bibitem[{{Metcalf} {et~al.}(2006){Metcalf}, {Leka}, {Barnes}, {Lites},
  {Georgoulis}, {Pevtsov}, {Balasubramaniam}, {Gary}, {Jing}, {Li}, {Liu},
  {Wang}, {Abramenko}, {Yurchyshyn}, \& {Moon}}]{Metcalf:2006}
{Metcalf}, T.~R., {Leka}, K.~D., {Barnes}, G., {et~al.} 2006, Sol. Phys., 237,
  267

\bibitem[{{Molodensky}(1969)}]{Molodenskii69}
{Molodensky}, M.~M. 1969, Soviet Astron.-AJ, 12, 585

\bibitem[{{Pietarila} {et~al.}(2012){Pietarila}, {Bertello}, {Harvey}, \&
  {Pevtsov}}]{Pietarila:2012}
{Pietarila}, A., {Bertello}, D., {Harvey}, F., \& {Pevtsov}, A. 2012, Sol.
  Phys.

\bibitem[{{R{\'e}gnier} \& {Amari}(2004)}]{Regnier:2004}
{R{\'e}gnier}, S. \& {Amari}, T. 2004, 425, 345

\bibitem[{{R{\'e}gnier} {et~al.}(2002){R{\'e}gnier}, {Amari}, \&
  {Kersal{\'e}}}]{Regnier:2002}
{R{\'e}gnier}, S., {Amari}, T., \& {Kersal{\'e}}, E. 2002, A\&A, 392, 1119

\bibitem[{{R{\'e}gnier} \& {Priest}(2007{\natexlab{a}})}]{Regnier:2007a}
{R{\'e}gnier}, S. \& {Priest}, E.~R. 2007{\natexlab{a}}, A\&A, 468, 701

\bibitem[{{R{\'e}gnier} \& {Priest}(2007{\natexlab{b}})}]{Regnier}
{R{\'e}gnier}, S. \& {Priest}, E.~R. 2007{\natexlab{b}}, \apjl, 669, L53

\bibitem[{{Ronan} {et~al.}(1987){Ronan}, {Mickey}, \& {Orrall}}]{Ronan:1987}
{Ronan}, R.~S., {Mickey}, D.~L., \& {Orrall}, F.~Q. 1987, Sol. Phys., 113, 353

\bibitem[{{Sandman} \& {Aschwanden}(2011)}]{Sandman:2011}
{Sandman}, A.~W. \& {Aschwanden}, M.~J. 2011, solphys, 270, 503

\bibitem[{{Schou} {et~al.}(2012){Schou}, {Scherrer}, {Bush}, {Wachter},
  {Couvidat}, {Rabello-Soares}, {Bogart}, {Hoeksema}, {Liu}, {Duvall}, {Akin},
  {Allard}, {Miles}, {Rairden}, {Shine}, {Tarbell}, {Title}, {Wolfson},
  {Elmore}, {Norton}, \& {Tomczyk}}]{Schou:2012}
{Schou}, J., {Scherrer}, P.~H., {Bush}, R.~I., {et~al.} 2012, Sol. Phys., 275,
  229

\bibitem[{{Schrijver}(2009)}]{Schrijver:2009a}
{Schrijver}, C.~J. 2009, Adv. Space Res., 43, 739

\bibitem[{{Schrijver} {et~al.}(2006){Schrijver}, {Derosa}, {Metcalf}, {Liu},
  {McTiernan}, {R{\'e}gnier}, {Valori}, {Wheatland}, \&
  {Wiegelmann}}]{Schrijver06}
{Schrijver}, C.~J., {Derosa}, M.~L., {Metcalf}, T.~R., {et~al.} 2006, Sol.
  Phys., 235, 161

\bibitem[{{Schrijver} \& {Title}(2011)}]{Schrijver:2011}
{Schrijver}, C.~J. \& {Title}, A.~M. 2011, J. Geophys. Res., 116, A04108

\bibitem[{{Skumanich} \& {Lites}(1987)}]{Skumanich:1987}
{Skumanich}, A. \& {Lites}, B.~W. 1987, \apj, 322, 473

\bibitem[{{Tadesse} {et~al.}(2009){Tadesse}, {Wiegelmann}, \&
  {Inhester}}]{tilaye09}
{Tadesse}, T., {Wiegelmann}, T., \& {Inhester}, B. 2009, A\&A, 508, 421

\bibitem[{{Tadesse} {et~al.}(2011){Tadesse}, {Wiegelmann}, {Inhester}, \&
  {Pevtsov}}]{Tilaye:2010}
{Tadesse}, T., {Wiegelmann}, T., {Inhester}, B., \& {Pevtsov}, A. 2011, A\&A,
  527, A30+

\bibitem[{{Tadesse} {et~al.}(2012{\natexlab{a}}){Tadesse}, {Wiegelmann},
  {Inhester}, \& {Pevtsov}}]{Tilaye:2012}
{Tadesse}, T., {Wiegelmann}, T., {Inhester}, B., \& {Pevtsov}, A.
  2012{\natexlab{a}}, Sol. Phys., 60

\bibitem[{{Tadesse} {et~al.}(2012{\natexlab{b}}){Tadesse}, {Wiegelmann},
  {Inhester}, \& {Pevtsov}}]{Tilaye:2012a}
{Tadesse}, T., {Wiegelmann}, T., {Inhester}, B., \& {Pevtsov}, A.
  2012{\natexlab{b}}, Sol. Phys., 277, 119

\bibitem[{{Thalmann} {et~al.}(2012){Thalmann}, {Pietarila}, {Sun}, \&
  {Wiegelmann}}]{Thalmann:2012}
{Thalmann}, J.~K., {Pietarila}, A., {Sun}, X., \& {Wiegelmann}, T. 2012,
  Astronomical J.

\bibitem[{{Thalmann} {et~al.}(2008){Thalmann}, {Wiegelmann}, \&
  {Raouafi}}]{Thalmann}
{Thalmann}, J.~K., {Wiegelmann}, T., \& {Raouafi}, N.-E. 2008, \aap, 488, L71

\bibitem[{{Turmon} {et~al.}(2010){Turmon}, {Jones}, {Malanushenko}, \&
  {Pap}}]{Turmon:2010}
{Turmon}, M., {Jones}, H.~P., {Malanushenko}, O.~V., \& {Pap}, J.~M. 2010, Sol.
  Phys., 262, 277

\bibitem[{{Unno}(1956)}]{Unno:1956}
{Unno}, W. 1956, Publ. Astron. Soc. Japan, 8, 108

\bibitem[{{Valori} {et~al.}(2012){Valori}, {Green}, {D{\'e}moulin}, {Vargas
  Dom{\'{\i}}nguez}, {van Driel-Gesztelyi}, {Wallace}, {Baker}, \&
  {Fuhrmann}}]{Valori:2012}
{Valori}, G., {Green}, L.~M., {D{\'e}moulin}, P., {et~al.} 2012, Sol. Phys.,
  278, 73

\bibitem[{{Valori} {et~al.}(2005){Valori}, {Kliem}, \& {Keppens}}]{valori05}
{Valori}, G., {Kliem}, B., \& {Keppens}, R. 2005, \aap, 433, 335

\bibitem[{{Wheatland}(2004)}]{Wheatland04}
{Wheatland}, M.~S. 2004, \solphys, 222, 247

\bibitem[{{Wheatland} \& {Leka}(2011)}]{Wheatland:2011}
{Wheatland}, M.~S. \& {Leka}, K.~D. 2011, ApJ, 728, 112

\bibitem[{{Wheatland} \& {R{\'e}gnier}(2009)}]{Wheatland:2009}
{Wheatland}, M.~S. \& {R{\'e}gnier}, S. 2009, \apjl, 700, L88

\bibitem[{{Wheatland} {et~al.}(2000){Wheatland}, {Sturrock}, \&
  {Roumeliotis}}]{Wheatland00}
{Wheatland}, M.~S., {Sturrock}, P.~A., \& {Roumeliotis}, G. 2000, ApJ, 540,
  1150

\bibitem[{{Wiegelmann}(2004)}]{Wiegelmann04}
{Wiegelmann}, T. 2004, Sol. Phys., 219, 87

\bibitem[{{Wiegelmann}(2007)}]{Wiegelmann07}
{Wiegelmann}, T. 2007, Sol. Phys., 240, 227

\bibitem[{{Wiegelmann} \& {Inhester}(2010)}]{Wiegelmann10}
{Wiegelmann}, T. \& {Inhester}, B. 2010, A\&A, 516, A107+

\bibitem[{{Wiegelmann} {et~al.}(2006){Wiegelmann}, {Inhester}, \&
  {Sakurai}}]{Wiegelmann06sak}
{Wiegelmann}, T., {Inhester}, B., \& {Sakurai}, T. 2006, Sol. Phys., 233, 215

\bibitem[{{Wiegelmann} {et~al.}(2012){Wiegelmann}, {Thalmann}, {Inhester},
  {Tadesse}, {Sun}, \& {Hoeksema}}]{Wiegelmann:2012}
{Wiegelmann}, T., {Thalmann}, J.~K., {Inhester}, B., {et~al.} 2012, Sol. Phys.,
  67

\end{thebibliography}

\end{document}